\documentclass[11pt,twoside]{article}

\usepackage{asp2004}
\usepackage{lscape}
\usepackage{graphicx}

\newcommand{\msun}{M_{\odot}}

\newcommand{\ltsim}{\protect\raisebox{-0.5ex}{$\:\stackrel{\textstyle <}
        {\sim}\:$}}
\newcommand{\gtsim}{\protect\raisebox{-0.5ex}{$\:\stackrel{\textstyle >}
        {\sim}\:$}}
\renewcommand{\nat}{Nature}

\begin{document}
\title{Massive Star Formation: A Tale of Two Theories}   
\author{Mark R. Krumholz\footnotemark}   
\affil{Princeton University}    

\begin{abstract} 
The physical mechanism that allows massive stars to form is a major
unsolved problem in astrophysics. Stars with masses $\gtsim 20$ $\msun$ reach
the main sequence while still embedded in their natal clouds, and the
immense radiation output they generate once fusion begins can exert a force
stronger than gravity on the dust and gas around them. They also
produce huge Lyman continuum luminosities, which can ionize and
potentially unbind their parent clouds. This makes massive star
formation a more daunting problem than the formation of low mass
stars. In this review I present the current state of the field, and
discuss the two primary approaches to massive star formation. One
holds that the most massive stars form by direct collisions between
lower mass stars and their disks. The other approach is to see if the
radiation barrier can be overcome by improved treatment of the
radiation-hydrodynamic accretion process. I discuss the theoretical
background to each model, the observational predictions that can be
used to test them, and the substantial parts of the problem that
neither theory has fully addressed.
\end{abstract}



\footnotetext{Hubble Fellow}

\section{Introduction}

Observations indicate that stellar initial mass function (IMF) is an
unbroken power law out to masses of about $150$ $\msun$
\citep{elmegreen00b, weidner04, figer05, oey05}, and there is no evidence for variation of either the
limit or the index of the mass function with metallicity or other
properties of the star-forming environment \citep{massey98}. Why the
mass limit for stars is so high, what physics sets it, and why the
mass spectrum seems to be universal are major unsolved problems in
astrophysics. Their solution requires a model for how massive stars
form, which at present is lacking due to both observational and theoretical challenges.

Massive stars form in the densest regions
within molecular clouds. We detect these massive star-forming clumps
as infrared dark clouds \citep[e.g.][]{rathborne05} or as millimeter
sources \citep[e.g.][]{plume97,shirley03}. The clumps have extremely high column
densities and velocity dispersions ($\Sigma \sim 1$ g cm$^{-2}$,
$\sigma \sim 4$ km s$^{-1}$ on scales of $\ltsim 1$ pc), and appear to
be approximately virialized. However, the structures within the
clumps that are the progenitors of single massive stars or
small-multiple systems are only now becoming accessible to
observations \citep{reid05, beuther05b}. Observations continue to
improve, but are hampered by large distances, heavy obscuration, and
confusion due to high densities.

On the theoretical side the problem is perhaps even more difficult.
Stars with masses $\gtsim 20$ $\msun$ have short
Kelvin-Helmholtz times that enable them to reach the main sequence
while still accreting from their natal clouds \citep{shu87}.
The resulting nuclear
burning produces a huge luminosity and a correspondingly large
radiation pressure force on dust grains suspended in the gas
surrounding the star. Early spherically symmetric calculations found
that the radiation force becomes stronger than gravity, and
sufficient to halt further accretion, once a star reaches a mass of
roughly $20-40$ $\msun$ \citep{kahn74, wolfire87} for typical Galactic
metallicities. More recent work has loosened this constraint by
considering the effect of an accretion disk. Disks concentrate the
incoming gas into a smaller solid angle, while shadowing most of it
from direct exposure to starlight \citep{nakano89, nakano95,
jijina96}. Cylindrically symmetric numerical simulations with disks
find that they allow accretion to continue up to just over $40$ $\msun$
before radiation pressure reverses the inflow
\citep{yorke02}. 

Ionization from a massive star presents a second problem to be
overcome. The escape speed in a massive star-forming core is
considerably smaller than the sound speed of $10$ km
s$^{-1}$ in ionized gas, so if a star is able to ionize its parent core into an HII
region, the core will be unbound
and accretion will stop \citep{larson71, yorke77}. Only if the
ionization is quenched near the stellar surface, where the escape
speed is larger than the sound speed, can accretion
continue. Early work on the problem of massive star formation
found that ionization was the dominant mechanism in setting an upper
mass limit on stars, although the later realization
that dust will absorb much of the ionizing radiation shifted
theoretical attention more towards the effects of radiation pressure.

Today, there are two dominant models of massive star formation. In
\S~\ref{compacc}, I present the competitive accretion model, in which
stars are born small and grow by accretion of unbound gas and
by collisions. In \S~\ref{radhydro}, I discuss the turbulent radiation-hydrodynamic model, which suggests that massive stars form from
massive, turbulent cores, and that neither radiation pressure nor
ionization prevents accretion onto a massive star. Finally, I discuss the missing pieces of the
picture that neither model is yet able to supply in \S~\ref{missing},
and summarize the state of the field and future prospects in
\S~\ref{conclusions}

\section{Competitive Accretion}
\label{compacc}

\subsection{The Model}

The competitive accretion model for massive star formation begins with
the premise that all stars are born small, with an initial mass
ranging from as much as $\sim 0.5$ $\msun$ \citep{bonnell04} to as
little as $\sim 3$ Jupiter masses \citep{bate05}, depending on the
particular variant of the theory. These ``seeds'' are born in a dense
molecular clump, and they immediately begin accreting gas to which they were not initially bound. Stars near the center of the clump are immersed in the highest
density, lowest velocity dispersion gas, and accrete most rapidly
\citep{bonnell01a, bonnell01b}. The clump is globally unstable to
collapse, and it contracts to stellar densities of $\sim 10^6-10^8$
pc$^{-3}$. At this point low mass stars begin to merge, either through
direct collisions \citep{bonnell98}, or because gas drag and
continuing accretion of low angular momentum gas causes binary
systems to inspiral \citep{bonnell05}. For example, a simulation
of a $1000$ $\msun$ clump with a radius of 0.5 pc by
\citet{bonnell03} produces a nearly $30$ $\msun$ binary system whose
members approach within $\sim 20$ AU of one another. In the
simulation gravity is softened on scales of 160 AU, so it is unclear
how the system would really evolve. However, \citet{bonnell05} argue
that it would likely merge, leading to the formation of a $20-30$
$\msun$ star.

One particularly appealing feature of the merger scenario is that it
provides a natural explanation for the observation that O and B stars
form solely (or almost solely) in rich clusters \citep{lada03} that
are strongly mass segregated \citep{hillenbrand98}. Since the rates of
competitive accretion and mergers are highest in cluster centers, and
both processes can only occur in clusters, this model qualitatively
reproduces the observations automatically. The model also naturally produces a high proportion of close, massive binaries, since for every binary that merges there are several more that come close \citep{bonnell05, pinsonneault06}.

Most work on competitive accretion to date uses no physics beyond
hydrodynamics and gravity. \citet{dale05} make a preliminary effort to
include ionization effects, but they focus more on the scale of
clusters than on individual stars, so their simulations do not have the
resolution to study how photoionization might affect accretion onto a
single star. No competitive accretion model to date includes either
magnetic fields or radiation pressure. The latter omission is
particularly important, since it means there is no evidence that competitive accretion by itself resolves the radiation pressure problem -- only mergers
do that. Indeed, simulations of Bondi-Hoyle accretion with
radiation find that radiation
pressure halts accretion onto stars with masses
$\gtsim 8$ $\msun$ \citep{edgar04} -- although these results appear
questionable in light of the more realistic simulations we discuss
in \S~\ref{radhydro} If Edgar \& Clarke's results do hold, though, so
that radiation pressure limits Bondi-Hoyle accretion (but not
accretion from a core) onto a massive binary, then the only way for
massive stars to grow in a competitive accretion model is by direct
collisions, rather than drag-induced binary mergers. This requires
stellar densities of $10^8$ pc$^{-3}$, $\sim 3$ orders of magnitude
larger than any observed to date in the Galactic plane. 

\subsection{Observational Evidence}

There are several potential direct observational signatures of the
competitive accretion scenario. \citet{bally05} suggest two
approaches: collisions would produce both infrared flares lasting
years to centuries and explosive, poorly-collimated outflows. At present there is no
data set available where one could search for the flares. For the
outflows, there is one known example that roughly fits the description
that Bally \& Zinnecker propose (the OMC-1 outflow), but there has
been no detailed modeling of how the outflow from a collision would
actually appear, and, as we discuss below, it is unclear how common such poorly collimated
outflows are. A third direct test is to search for embedded clusters
with densities of $\sim 10^8$ pc$^{-3}$, which are a required
component of the competitive accretion picture. Such objects should be
short-lived and therefore rare, but their high column densities would
give them a distinct spectral shape that might be observable
with Spitzer, and should be easily observable by JWST or Sophia
(S. Chakrabarti \& C.~F. McKee, 2006, in preparation).

One can also use more indirect tests to look for evidence of
mergers, and here the competitive accretion picture runs into
considerable difficulty. If massive stars form via collisions, the
collision process should truncate their accretion disks or disrupt them
entirely. The collision itself may give rise to a fat torus, but is
unlikely to produce a thin disk \citep{bally05}. Thus, the collisional
formation model predicts that massive stars should not have disks
hundreds of AU in size, as are observed for low mass stars. However,
there are now at least two known examples of massive stars with such
large disks \citep{jiang05, patel05}.
Since thin disks are probably required to create well-collimated MHD
outflows, the collision scenario also predicts that massive protostars
should lack well-collimated outflows. However, interferometric
observations of young massive stars reveal that outflows for stars as
massive as early B usually are well-collimated \citep[][and references
therein]{beuther05}. Position-velocity diagrams \citep{beuther04} and
near-IR images \citep{davis04} of outflows, as well as correlations
between outflow momentum and luminosity of the driving star
\citep{richer00}, also point to a common driving mechanism for low
mass and high mass protostellar outflows, inconsistent with the
competitive accretion / collision scenario.

\subsection{Theoretical Difficulties}
\label{compaccproblem}

The apparent conflict between competitive accretion models and
observations has led to theoretical reconsideration of the
problem. For competitive accretion to be effective a small ``seed''
protostar in a molecular clump must be able to accrete its own mass or more
within a dynamical time of its parent clump. The process by which the
protostar gathers gas from the clump is Bondi-Hoyle accretion in a
turbulent medium, a process for which \citet{krumholz05b, krumholz06a}
give a general theory supported by simulations. Using this result,
together with an analysis of the possibility that protostars might
gain mass by capturing other cores in their parent clump,
\citet{krumholz05e} determine what properties a star-forming molecular
clump must have for competitive accretion within it to be
effective. They show that competitive accretion only works in clumps
with $\alpha_{\rm vir}^2 M\ltsim 10$ $\msun$, where $M$ is the clump
mass and $\alpha_{\rm vir}$ is the clump virial parameter
\citep{bertoldi92, fiege00b}, roughly its ratio of kinetic energy to
gravitational potential energy. For observed star-forming clumps,
$\alpha_{\rm vir}\sim 1$ and $M\sim 1000$ $\msun$, so competitive
accretion will not operate. It occurs in simulations only because
the regions simulated have smaller virial parameters and masses than
observed regions. In some cases the virial parameters are too small to
begin with \citep{bonnell01a, bonnell01b}, and in others the virial
parameters start near unity, but decay of turbulence quickly reduces
them to smaller values \citep{bonnell04, bate02a, bate02b,
bate03}. The results of \citet{krumholz05e} strongly suggest that
competitive accretion plus mergers cannot be the mechanism by which
massive stars form.

\section{Turbulent Radiation Hydrodynamic Models}
\label{radhydro}

Given the difficulties with the competitive accretion scenario, one
must ask whether it might be possible to form massive stars in roughly
the same way as low mass stars, via collapse from a coherent core
and disk accretion. Such a scenario must overcome three
serious challenges: one requires a plausible model for the origin and
structure of massive cores, a method to allow accretion to occur
despite radiation pressure feedback, and an explanation for why
ionization does not destroy the protostellar core before the massive
star is fully assembled.

\subsection{Massive Cores}

Theoretical arguments predict
that fragmentation in a turbulent medium produces a spectrum of bound
fragment masses that resembles the stellar IMF \citep{padoan02}. If
these arguments are correct, then massive cores are simply the tail of
the distribution of core masses. Simulations of fragmentation in a
turbulent medium do roughly concur with theoretical models
\citep{li04}, and observations also support the idea that cores have a
mass distribution that parallels the stellar IMF, and that
cores with masses $\gg \msun$ exist \citep[e.g][]{motte98, testi98,
johnstone01, reid05, beuther05b}. Thus, massive cores may naturally
arise from turbulent fragmentation.

Massive cores, however, must be structured somewhat differently than
the cores that give rise to low mass stars. The thermal Jeans mass in
star forming regions is $\sim 1$ $\msun$, so massive
cores cannot be supported primarily by thermal pressure. Instead, they
must be turbulent. \citet{mckee03} present a self-similar model
of massive, turbulent cores that are in rough pressure balance with
the high pressure environments they form. This gives them surface
densities $\sim 1$ g cm$^{-2}$ and pressures $P/k \sim 10^8$ K
cm$^{-3}$, much larger than the mean column density and pressure in
GMCs. These high pressures cause the cores to be extremely compact,
with radii $\ltsim 0.1$ pc, and the correspondingly high density
produces accretion rates of $\sim 10^{-3}$ $\msun$ yr$^{-1}$ onto
embedded stars, allowing massive stars to form in $\sim 10^5$ yr.

One important question for models of massive cores is whether they
will produce one or a few massive stars, or fragment
to produce numerous low mass stars instead. \citet{dobbs05} simulate
centrally condensed turbulent cores with structures that follow the
\citet{mckee03} model, and find that they form many low mass stars
rather than a single massive star. However, their simulations do not include radiation. \citet{krumholz06b} perform similar
simulations including radiative transfer, and find that the
combination of high accretion luminosity and high optical depth that
occur in high-density cores produce rapid heating that inhibits
fragmentation. Of course the massive core models
used by both Dobbs et al. and Krumholz et al. are highly idealized, so
the question of to what extent real massive cores fragment remains open.

\subsection{Accretion with Radiation Pressure}

\subsubsection{The Flashlight Effect}

Once a massive protostar reaches $\sim 15$ $\msun$, the pressure
exerted by its radiation field will begin to have a significant effect
on the accretion flow. Two dimensional radiation-hydrodynamic
simulations by \citet{yorke02}
find that, once the radiation field becomes significant, the
it begins to reverse inflow along the
poles. Accretion continues through an accretion disk in the equatorial
plane, and the disk serves to collimate the radiation field and beam
it preferentially in the polar direction. This collimation is called
the flashlight effect. However, in Yorke \& Sonnhalter's simulations
this is not enough to allow very massive stars to form. As the
protostellar mass and luminosity increase, inflow
stops over a wider and wider range of angles about the
pole. Eventually, no more material is able to fall onto the disk, and
soon thereafter the radiation field disperses the disk itself. Yorke
\& Sonnhalter find a maximum final mass of the star of $\approx 20$
$\msun$ in simulations with gray radiation, and $\approx 40$ $\msun$
in simulations with a multi-frequency treatment of the
radiation field. The difference in outcome is likely due to
enhancement of the flashlight effect by the more realistic
multi-frequency radiation model.

\begin{figure}[ht!]
\centerline{\includegraphics[scale=0.8]{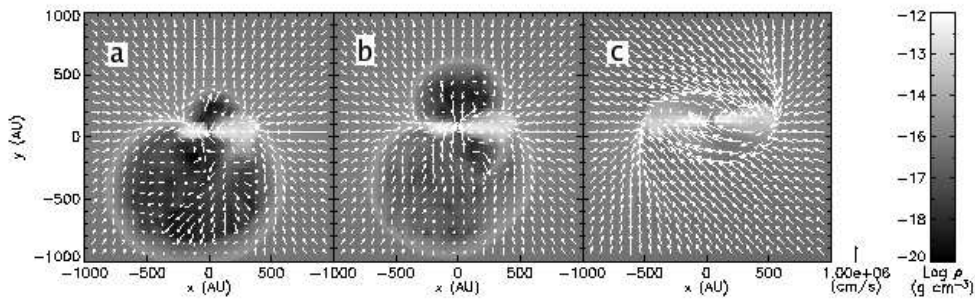}}
\caption{
\label{bubblev}
The plot shows a simulation of the collapse of a $100$ $\msun$ core by
\citet{krumholz06b}. Each panel is a slice in the XZ plane at a
different time, showing the density (grayscale) and velocity
(arrows). The times of the three slices are (a) $1.5\times 10^4$, (b)
$1.65\times 10^4$, and (c) $2.0\times 10^4$ yrs, and the stellar
masses at those times are $21.3$ $\msun$, $22.4$ $\msun$, and $25.7$ $\msun$.
}
\end{figure}

\begin{figure}
\centerline{\includegraphics[scale=0.4]{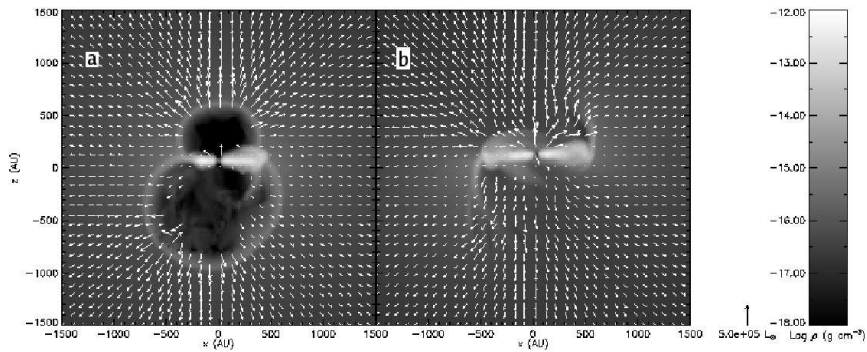}}
\caption{
\label{bubblef}
The plot shows a simulation by \citet{krumholz06b}. Panels (a) and (b)
are the same as Figure \ref{bubblev}, panels (b) and (c), but the
arrows show radiation flux rather than velocity. The flux vectors are
multiplied by $4\pi r^2$, where $r$ is the distance from the central
star. For clarity, flux vectors from inside the optically thin bubble
interior in panel (a) have been omitted.
}
\end{figure}

More recent three-dimensional radiation-hydrodynamic simulations,
however, demonstrate a qualitatively new effect that allows accretion to
higher masses that two-dimensional work suggests. \citet[][and 2006, in
preparation]{krumholz06b} find that at masses $\ltsim 17$ $\msun$, the
radiation field is too weak to reverse the inflow, and massive
cores evolve much as \citet{yorke02} find. At larger masses, the
radiation field begins to inhibit accretion along the poles, driving
bubbles into the accreting gas. However, the three-dimensional
simulations show that bubbles grow asymmetrically due to an
instability that is suppressed in Yorke \& Sonnhalter's
two-dimensional, single quadrant (i.e. assuming symmetry about the
$xy$ plane) simulations. Figure \ref{bubblev}a shows this
effect. Since the gas is extremely optically thick to stellar
radiation, the bubbles are able to collimate the radiation field and
beam it preferentially in the polar direction, as shown in Figure
\ref{bubblef}a. At the time shown in the Figure, the flux of radiation
in polar direction at the edge of the bubble is larger than the flux
in the equatorial plane by more than an order of magnitude. The strong
flux in the polar direction deflects gas that reaches the bubble walls
to the side. As the velocity field in Figure \ref{bubblev}a shows, it
then travels along the bubble wall and falls onto the disk, where it is
shielded from the effects of radiation by the disk's high optical
depth. The gas then accretes onto the star.

Eventually the instability becomes violent enough for the bubbles to
start collapsing (Figure \ref{bubblev}b), leaving behind remnant
bubble walls (Figure \ref{bubblev}c). These dense walls serve to
collimate the radiation and shield the gas from it, as shown in Figure
\ref{bubblef}b, allowing it to reach the accretion disk and then the
star. The collapse is in essence a radiation Rayleigh-Taylor
instability, caused by the inability of radiation, a light fluid, to
hold up the heavy gas. Simulations to date have reached masses of
$\approx 34$ $\msun$ onto the star, with $5-10$ $\msun$ in the
disk. Thus far there are no sign of accretion being reversed, and the
simulations are continuing as of this writing. Note that these
calculations use gray radiative transfer, a case for which
\citet{yorke02} found a limit of $\approx 20$ $\msun$; Yorke \&
Sonnhalter's results suggest collimation of the radiation field would
be even more effective, and accretion correspondingly easier, with
multi-frequency radiation.

\subsubsection{Protostellar Outflows}

The simulations discussed above all ignore the presence of
protostellar outflows. However, \citet{krumholz05a} point out
that outflows can also have a strong effect on the radiation
field. Outflows from massive stars are launched from close to the
star, where radiation heats the gas to the point where all the
dust sublimes. As a result, outflows are dust-free and very
optically thin when they are lunched. Because outflows leave the
vicinity of the star at high speeds ($\gtsim 500$ km s$^{-1}$), there
is no time for dust within the outflow cavity to re-form before the
gas is well outside the collapsing core. Because the core around
it is very optically thick, the outflow cavity collimates radiation
and carries it away very efficiently. It effectively
becomes a pressure-release valve for the radiation. Monte Carlo
radiative transfer calculations show that, for outflow cavities
similar to those observed from massive protostellar outflows, the
presence of a cavity can reduce the radiation pressure force
on infalling gas by an order of magnitude. This can shift the inflow
from a regime where radiation pressure is stronger than gravity to one
where it is weaker. \citet{krumholz05b} show that, even for a 50
$\msun$ star embedded in a 50 $\msun$ envelope, an outflow cavity
would make radiation pressure weaker than gravity over a quarter
of the solid angle onto the star.

It is unclear how radiation collimation by outflows will interact with
the radiation bubbles and Rayleigh-Taylor instability that occur in
simulations where an outflow is not present. However, the overall
conclusion one may draw from both effects is that, in an optically
thick core, it is very easy to collimate radiation. Collimation
reduces radiation pressure over much of the available solid angle, and
allows accretion to continue to higher masses than naive estimates
suggest. Radiation pressure is not a
significant barrier to accretion.

\subsection{Ionization}

The third puzzle to solve in an accretion mechanism for massive star
formation is ionization. For spherically symmetric accretion,
\citet{walmsley95} shows that accretion above a critical rate
\begin{equation}
\dot{M}_{\rm crit} = \sqrt{\frac{4\pi G M S m_H^2}{\alpha^{(2)}}}
\approx 2\times 10^{-5}
\left(\frac{M}{10\;\msun}\right)^{1/2}
\left(\frac{S}{10^{49}\mbox{ s}^{-1}}\right)^{1/2} 
\;\msun\mbox{ yr}^{-1},
\end{equation}
where $M$ is the stellar mass, $S$ is the ionizing luminosity (in
photons s$^{-1}$), and $\alpha^{(2)}$ is the recombination coefficient
to excited levels of hydrogen, will trap all ionizing photons near the
stellar surface. Since estimated accretion rates for massive stars are
much higher than this, if accretion is spherically symmetric then the
star will be unable to ionize its parent core.

Observations support the idea that HII regions can be confined near their source stars for long periods, and that they are therefore not able to stop accretion. The argument comes from statistics:  ultracompact HII regions (roughly those $\ltsim 0.1$ pc in size) have dynamical times $\sim 10^3$ yr, but a census of the number of HII regions versus their size implies that ultracompact HII regions must
survive for times closer to $\sim 10^5$ yr  \citep{wood89, kurtz94}. An extended phase during which
HII regions are confined by accretion, and which lasts for a time comparable to the star formation time ($\sim 10^5$ yr), is consistent with the data, while the idea that HII regions expand dynamically and halt accretion is not. In addition, observations of inflow signatures in ionized gas in some systems \citep{sollins05} provide direct evidence that accretion can continue even after the formation of an HII region.

The exact mechanism by which HII regions are confined is still
uncertain. \citet{keto02,keto03} presents spherically symmetric
hydrodynamic accretion models with ionizing radiation. In the models,
when the ionizing luminosity is low, accretion traps all ionizing
photons at the stellar surface and there is no HII region. As the
stellar mass and ionizing luminosity increase, the HII region is able
to lift off the stellar surface, but it remains trapped in a region
where the thermal pressure of the ionized gas is insufficient to
escape. Accretion continues through the ionization front for a long
time, but eventually the ionizing luminosity rises high enough for the
HII region to expand outward and reach the point where it halts
further accretion. While this model seems to be consistent with the
observational data, there are two possible problems. First, it is spherically symmetric, while massive stars form primarily in very turbulent regions. Second, the long trapped HII region phase requires that the ionizing luminosity and the accretion rate rise together. Ionizing luminosity rises sharply with stellar mass, and in Keto's models (which assume Bondi accretion) so does the accretion rate. However, if the accretion rate were a weaker function of mass, as is expected for more realistic core models \citep[e.g.][]{mckee03}, then it is unclear the confinement would work.

\citet{xie96} offer another possibility: HII regions may be confined by turbulent pressure, which far exceeds thermal pressure in the dense regions where massive stars form. However, Xie et al.'s model is purely analytic, and it is unclear if turbulent confinement of ionization can work in reality. Recent simulations by \citet{dale05} of collapsing regions suggest that it will not, because ionizing radiation will escape through low-density regions of the turbulent flow. In Dale et al.'s simulations the turbulent pressure far lower than it should be due to turbulent decay (\S~\ref{compaccproblem}), but the results suggest at a minimum that turbulent confinement of HII regions needs further study.
\citet{tan03} ofter the alternative model that HII regions may be
confined by the dense outflows of massive stars. In this picture, the
ionized region is confined to the dense walls of an outflow cavity
that has been largely evacuated of gas by magnetic fields. They find
that the ionization stays confined near the star as long as the star
for B star or later ionizing luminosities. However, it is unclear what this model predicts will happen for more massive stars.

Regardless of which model for trapping is correct, it is important to note that \textit{some} mechanism for confining HII regions for many dynamical times seems to be required by the observational data. All the proposed explanations thus far involve accretion, winds, or turbulence in some form. There is no plausible solution for the long lifetimes of HII regions in the competitive accretion picture, which lacks these elements and generally predicts gas (as opposed to collisional) accretion rates onto massive stars that are too low to confine ionization \citep{dale05, dobbs05}. This is another serious argument against competitive accretion.

\section{Missing Pieces}
\label{missing}

Thus far, I have argued that the turbulent radiation hydrodynamic
model provides a good solution to the problem of massive star
formation. However, there are several elements of the problem for
which neither that model nor the competitive accretion model have made
much progress.

\subsection{Magnetic Fields}

None of the simulations of massive star formation performed to date
have included magnetic fields, and analytic models have included them
only in the most cursory fashion. It is unclear how serious an omission
this is. The dynamical importance of the magnetic field is determined
by the mass to magnetic flux ratio, $M/\Phi$. For a given flux there
is a maximum mass $M_{\Phi}$ that flux can support against gravitational
collapse, so for a cloud of mass $M$ it is natural to define the magnetic
support parameter $\lambda = M/M_{\Phi}$. Values of $\lambda>1$ are termed supercritical,
and correspond to configurations where magnetic support cannot
prevent the cloud from collapsing dynamically; in the subcritical
regime, $\lambda < 1$, magnetic support can prevent collapse. If
typical massive star forming cores are subcritical or critical, then
omission of magnetic fields is a serious error.

In principle one can determine $\lambda$ directly from
observations. In practice, however, magnetic fields are extremely
difficult to detect even in low-mass star forming regions, which are
generally closer and suffer much less from confusion and extinction
than massive regions. \citet{crutcher06} reviews the observations of
magnetic fields in massive star-forming regions that exist, and
concludes that $\lambda\approx 1$ is typical, indicating the
magnetic effects are significant.

However, this conclusion is plagued by large systematic
uncertainties. First, to determine $\lambda$ from observations, one
must assume a geometry for the cloud. Crutcher's conclusion assumes
that cores are two-dimensional disks. However, observations show that
cores are roughly triaxial \citep{jones01}, with ratios of long to
short axis of $\sim 2:1$. This would give $\lambda > 1$ for the vast
majority of observed regions. A second difficulty stems from
uncertainty in where within a core one is measuring a magnetic
field. The most common and reliable way to detect magnetic fields in
molecular gas is via Zeeman splitting in OH or
CN. However, both of these molecules are biased tracers of the mass,
due to excitation threshold effects and freeze-out onto dust grains
\citep{tafalla02}. It is unclear what systematic biases observing the field in these biased tracers might produce. Methods such as the Chandrasekhar-Fermi
effect, which are based on polarization of dust grains, are not
affected by freeze-out, but are affected by uncertainty as to where
along a line of sight a polarized signal is arising. It is not clear
whether these effects will systematically increase or decrease
$\lambda$. A third bias is that in many cases observations
do not detect a magnetic field at all, and at least some non-detections remain unpublished. If such regions are not properly included in statistical analyses, this can artificially raise estimates of $\lambda$ \citep{bourke01}. In summary,
observations are quite ambiguous as to whether magnetic fields
are dynamically significant in regions of massive star
formation. Ideally they should be included in models, but limitations
of algorithms have prevented their inclusion thus far.

\subsection{Masers}

Decades ago observations established that massive star forming
regions are often host to water, methanol, OH, and SiO masers. Although they were originally thought to arise from shocks at the edges of HII regions, high resolution
observations show that they are generally offset from HII regions, and
are more closely associated with infrared sources \citep{hofner96}. 
Masers are particularly useful because they provide spatial resolutions of milliarcseconds, far higher than than any other technique possible for deeply embedded sources. The
resolution in space and time is sufficiently high that multi-epoch observations can often detect proper motions of individual maser spots.

The primary difficulty with maser observations is that they are
difficult to interpret. Maser spots often show linear or arc-like
arrangements, which early observers interpreted as tracing edge-on
disks \citep{norris93}. This would have been interesting, because at
the time no disks around massive stars were known. Even today, it would provide us with a powerful tool for tracing the dynamics of massive accretion disks on very small scales. However, more recent work that has combined maser data with observations in other wavelengths provides little support to the disk hypothesis. In a few cases, such as Orion BN/KL \citep{greenhill03}, linear arrangements of maser spots are perpendicular to molecular outflows, as one would expect were masers tracing a disk. More often, however, maser spots are parallel to the direction of outflows \citep{debuizer03, debuizer05, debuizer06}, suggesting that they trace outflows rather than disks.
The primary lesson is that masers cannot be used as diagnostics of the massive star formation process without more complete models of how and where maser emission arises.

\begin{figure}[ht!]
\centerline{\includegraphics[scale=0.65]{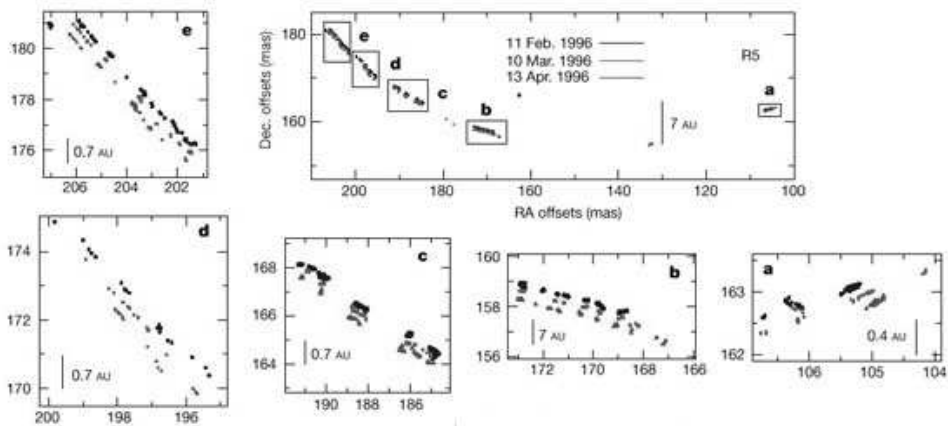}}
\caption{
\label{masercircle}
In the Figure, the dots show spots of water maser emission observed by \citet{torrelles01} in the Cepheus A star-forming region. Figure appears by the kind permission of the Nature Publishing Group.
}
\end{figure}

While difficult, making such models is likely to yield new insights
into the star formation process, particularly when applied to some of
the more unusual arrangements of masers that appear consistent with
neither a disk nor an outflow. For example, Figure \ref{masercircle}
reproduces Figure 1a of \citet{torrelles01}. The dots show the
positions of water maser spots observed in three epochs in the Cepheus
A region, a site of massive star formation. At each epoch the spots
fit a circle roughly 62 AU in radius around the same center to an accuracy of 0.1\%. The change in radius with time implies that the circle is expanding at 9 km s$^{-1}$. The best explanation for this geometry is that we are seeing a limb-brightened, expanding spherical shell, and there is no obvious way that either a disk or an ordinary bipolar outflow could explain the data. One intriguing possibility is that the masers could be tracing the wall of a radiation bubble, as seen in the simulations of \citet{krumholz06b}. The bubbles are quite spherical when they are at such small radii, the expansion velocities are roughly consistent with what is seen in the simulations, and the densities and temperatures in the bubble walls are roughly what would be needed to produce maser emission.

\subsection{The Stellar Mass Limit}

A final observational result that neither model has been able to incorporate or explain thus far is the existence of an upper mass limit to the stellar IMF. Statistical arguments applied to the Galaxy as a whole \citep{elmegreen00b,oey05} and direct star counts in individual massive clusters \citep{weidner04,figer05} both show that the IMF cannot continue to have a Salpeter slope out to arbitrarily high masses. Instead, there must be a fairly sharp turn-down at around $150$ $\msun$. It is difficult to see how such a cutoff could occur in the competitive accretion model. Collisions between point particles should be a scale-free process,  producing a featureless power-law distribution of masses. It is possible that the ``microphysics" of the collision process could provide a break in the power-law -- for example collisions between stars with a total mass above $150$ $\msun$ might lead directly to intermediate mass black holes rather than to stars that we could observe -- but there is at present no evidence to support this hypothesis.

The upper mass limit is not much easier to understand in the context of the turbulent radiation hydrodynamic model. One might think that the increasing strength of radiation pressure feedback with mass could produce a cutoff, but this explanation faces two serious objections. First, at masses $\gtsim 100$ $\msun$, stellar luminosity is almost directly proportional to mass, since at such masses stars are supported primarily by internal radiation pressure. Thus, the ratio of radiation pressure force to gravitational attraction does not change significantly for stars larger than $\sim 100$ $\msun$. Why, then, should be there be a change in the IMF at 150 $\msun$? A second problem with an explanation based on radiation pressure is the absence of evidence for a variation in the stellar IMF with metallicity. Since the strength of the radiation pressure force is directly proportional to the metallicity, if radiation pressure set the stellar mass limit then one would expect the high mass end of the IMF to change with metallicity, which should be observable as a change in IMF with Galactocentric radius. Such a change has not been observed \citep{massey98}. 

One possible way out would be if the mass limit is unrelated to the formation process, and is instead set by stellar stability. \citet{humphreys79,humphreys94} investigate the structure of very massive stars, and find that they are often pulsationally unstable. This instability can cause rapid mass loss, which might set a stellar upper mass limit. However, whether such a limit really exists, and if so whether it coincides with the observed mass limit, is at present unknown.

\section{Conclusions and Prospects}
\label{conclusions}

Our knowledge of the physical mechanism of massive star formation is still quite limited, as evidenced by the fact that for the last decade there have been two very different models for it that observations could not definitively distinguish. However, theoretical and observational work over the last year or two have advanced the field to the point where we can begin to decide between the models. Observations of disks and outflows from young massive stars point to accretion from a core rather than collision as the mechanism by which massive stars form, and theoretical work strongly suggests that competitive accretion does not operate in observed star-forming clouds, consonant with observations favoring accretion from cores. Moreover, the problem of how to make massive stars despite radiation feedback, one of the original motivations for the competitive accretion and collision model, seems to be receding. Recent simulations and theoretical work show that both radiation pressure force and ionization are much less effective at inhibiting accretion than had previously been assumed.

In the next decade, observations should be able to settle definitively
the issue by searching for more direct indicators of collision, such as very high column density embedded clusters and infrared flares from collisions. They also promise to give us a window into the massive star formation process on much smaller scales, where the effects of radiation pressure and ionization should be more obvious. Masers have started to provide data with high resolution in space and time, but interpreting maser data still requires much theoretical work. The next generation of millimeter interferometers, such as ALMA, will enable us to resolve disks around massive stars, and possibly to see dense shells of material shaped by protostellar radiation and outflows on $\ltsim 1000$ AU scales. These observations should be much easier to interpret.

On the theoretical side, progress will depend primarily on improving computational models, and should focus on four problems with the current generation of simulations. First, no simulation of massive star formation to date has included outflows. This is a major omission, since we know that outflows are present, and that they can have profound effects on the formation process. Outflows may also be responsible for driving turbulence in star-forming clumps, and should therefore be included in simulations of cluster formation to avoid the problem of unphysically small virial parameters identified by \citet{krumholz05e}.
Improving the computations from hydrodynamics to magnetohydrodynamics
is a second potential advance. The major difficulty here is knowing
what initial conditions to use, since the strength and geometry of the
magnetic field in regions of massive star formation is so poorly
known. Third, simulations could be improved by starting from larger
scales. Both competitive accretion and turbulent radiation
hydrodynamic simulations of massive star formation start with
extremely unrealistic initial conditions. A better approach would be
to simulate a cluster-forming clump $\sim 4000$ $\msun$ in size,
typical of the \citet{plume97} sample, follow the formation of a
massive core, and then simulate the subsequent collapse of the core at
high resolution using adaptive mesh refinement or adaptive smoothed
particle hydrodynamics.

A final area ripe for improvement is radiative transfer. Thus far simulations have either used multi-frequency radiation in two dimensions \citep{yorke02}, or gray radiation in three dimensions \citep{krumholz06b}. Since the simulations show that both multi-frequency and three-dimensional effects are important, it is critical to do three-dimensional multi-frequency radiative transfer simulations.
A natural outgrowth of this is modeling ionization, since in principle one can treat Lyman continuum photons as just another frequency group and then add a chemistry update step to recompute the ionization fraction after a radiation update. A final improvement to the radiation would be to use an approximation better than flux-limited diffusion, which may produce errors inside low optical-depth radiation bubbles. The major obstacle here is computational cost. Three-dimensional gray flux-limited diffusion calculations require months of supercomputer time on present computers, and improvements to the radiation physics without significant advances in processor or algorithmic speed would make the problem impossible to run.

Perhaps the best opportunities for progress now come not purely from
theory or from observation, but from work that makes detailed
comparisons of the two. Hopefully in the future more observers and
theorists will collaborate to post-process simulations so that they
can make definite comparisons to observations, and use the results of
those comparisons to refine theoretical models. In the next decade,
work of this sort should be able to provide us with at least the basic
outline of how massive stars form.

\acknowledgements I thank S.~C. Chakrabarti, C.~F. McKee, and
J.~C. Tan for helpful discussions. Support for this work was provided
by NASA through Hubble Fellowship grant \#HSF-HF-01186 awarded by the
Space Telescope Science Institute, which is operated by the
Association of Universities for Research in Astronomy, Inc., for NASA,
under contract NAS 5-26555.


\end{document}